\documentclass[aps,twocolumn,prl]{revtex4}
\usepackage{graphicx}
\usepackage{mathrsfs}
\usepackage{amsmath}

\begin{document}

\title{Polymer-disordered liquid crystals: Susceptibility to electric field}

\author{Lena M. Lopatina}
\altaffiliation{Current address: Los Alamos National Laboratory, Los Alamos, NM 87545}
\affiliation{Liquid Crystal Institute, Kent State University, Kent, Ohio 44242, USA}
\author{Jonathan V. Selinger}
\email{jselinge@kent.edu}
\affiliation{Liquid Crystal Institute, Kent State University, Kent, Ohio 44242, USA}

\date{August 12, 2013}

\begin{abstract}
When nematic liquid crystals are embedded in random polymer networks, the disordered environment disrupts the long-range order, producing a glassy state. If an electric field is applied, it induces large and fairly temperature-independent orientational order. To understand the experiments, we simulate a liquid crystal in a disordered polymer network, visualize the domain structure, and calculate the response to a field. Furthermore, using an Imry-Ma-like approach we predict the domain size and estimate the field-induced order. The simulations and analytic results agree with each other, and suggest how the materials can be optimized for electro-optic applications.
\end{abstract}

\maketitle

Over the last twenty years, liquid crystals in random environments have been studied as model systems for statistical mechanics under quenched disorder~\cite{crawford96,iannacchione04}.  Experiments have investigated liquid crystals confined to disordered polymer networks, aerogels, aerosils, and porous glasses~\cite{wu95,iannacchione96,zhou97,iannacchione97,bellini98,iannacchione98,bellini01,jin01,marinelli01,retsch02,park02,leheny03,iannacchione03,clegg03,roshi04,kutnjak04}.  Theoretical studies have shown that the quenched randomness of the environment provides a crucial disordering influence, beyond the randomness arising from thermal fluctuations, which can induce a glassy state all the way down to zero temperature~\cite{olmsted96,radzihovsky97,jacobsen99,feldman00}.  In this body of work, one issue that has generally \emph{not} been considered is the susceptibility of the glassy liquid-crystal state to a symmetry-breaking field.  The purpose of this paper is to point out that the susceptibility is an important scientific issue, which is also related to applications in electro-optic technology.  We calculate the susceptibility as a function of temperature and material properties, and show the parameters that must be optimized for applications.

For applications, many investigators have developed liquid-crystal displays and other electro-optic devices based on the Kerr effect, in which an electric field applied to an optically isotropic material induces orientational order and hence optical birefringence. This technology requires a Kerr coefficient that is large and approximately independent of temperature. One approach to achieve this goal is by using liquid-crystal blue phases~\cite{hisakado05,kitzerow06}, complex configurations of double-twist cylinders threaded by disclination lines~\cite{wright89}. Although blue phases are locally anisotropic, the anisotropy averages out on longer length scales, leading to zero birefringence. However, when an electric field $E$ is applied, it favors alignment of the molecules with respect to the field direction, thus inducing globally averaged orientational order and birefringence proportional to $E^2$~\cite{fukuda09,henrich10}. The main limitation of this approach is that blue phases are normally stable over only a very narrow temperature range. For that reason, substantial effort is being devoted to chemical synthesis~\cite{coles05} and polymer stabilization~\cite{kikuchi02} of blue phase materials.

As an alternative to blue phases, Yang and Yang have suggested using a nematic liquid crystal trapped within a disordered polymer network~\cite{yang11}. On short length scales the nematic phase has orientational order, but on long length scales the polymer disrupts the order, producing a glassy state with local nematic domains. Yang and Yang refer to the resulting structure as a ``polymer-stabilized isotropic phase,'' but it could equally well be called a ``polymer-disordered nematic.'' When an electric field is applied to this phase, it induces long-range orientational order of the local nematic domains, and hence the material becomes birefringent. For that reason, the material can function just as a blue phase, with a large Kerr effect, but without the need to stabilize it over a significant temperature range. Thus, it offers an excellent opportunity for electro-optic applications.

\begin{figure*}
\includegraphics[width=6.5in]{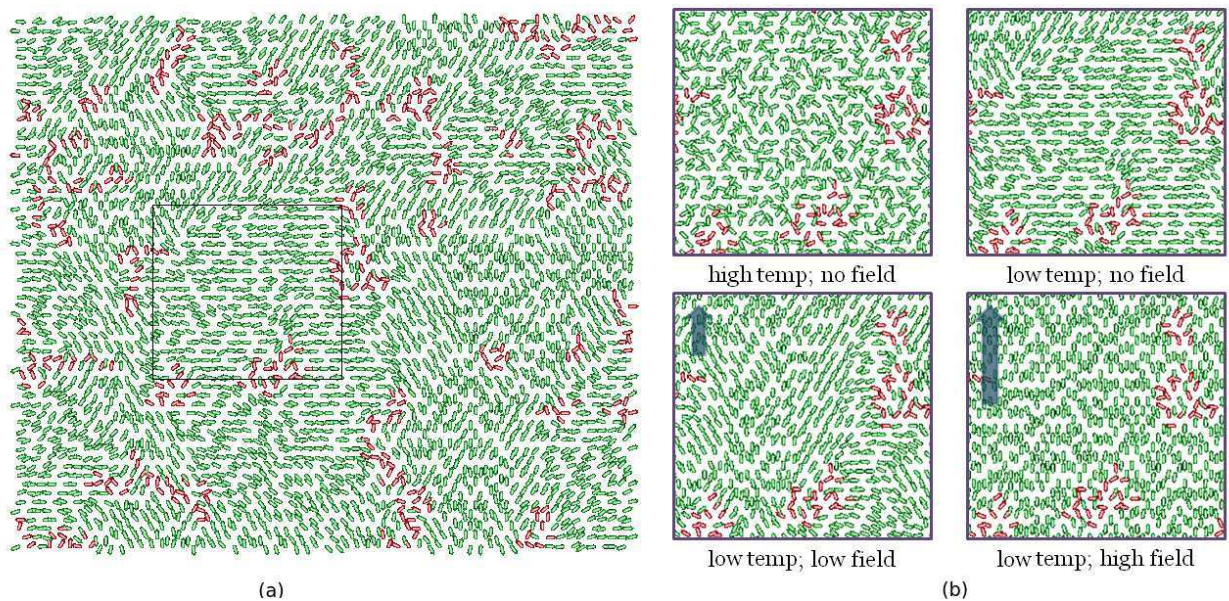}
\caption{Director profile and polymer orientations in the simulations. Green sites are liquid crystal; red sites are polymer. (a)~At low temperature the directors form domains with local orientational order, which cannot extend over a long range. (b) The same region of material at high and low temperature without electric field and under applied electric field.  At low field there is partial alignment of the liquid-crystal directors with electric field, and at high field the directors are completely aligned with the field.}
\label{fig_sim_temp}
\end{figure*}

In this paper, we develop a model for the susceptibility of a polymer-disordered nematic phase to an applied electric field, using Monte Carlo simulations and an Imry-Ma-like domain argument. For Monte Carlo simulations, the system is modeled by a two dimensional (2D) triangular lattice of unit vectors, as shown in Fig.~\ref{fig_sim_temp}(a). Each polymer is represented by a fixed chain of lattice sites, while the remaining sites represent liquid crystal molecules. The orientations of the polymer sites are fixed at the beginning of the simulation, and remain unchanged to mimic the effect of polymers that induce local random orientations on the neighboring liquid crystals, whose orientations are free to change in the Monte Carlo process. The Hamiltonian is
\begin{eqnarray}
\mathscr{H} &=& -J\sum_{\langle i_\mathrm{LC},j_\mathrm{LC}\rangle}{(\mathbf{\hat{n}}_i \cdot \mathbf{\hat{n}}_j)^2}
-H \sum_{\langle i_\mathrm{LC},j_\mathrm{P}\rangle}{ (\mathbf{\hat{n}}_i \cdot \mathbf{\hat{m}}_j)^2}\nonumber\\
&& - \sum_{i_\mathrm{LC}}{(\mathbf{E} \cdot \mathbf{\hat{n}}_i)^2}.
\label{hamiltonian_cs_3}
\end{eqnarray}
Here, the first term represents the interaction between neighboring liquid-crystal molecules at sites $i$ and $j$, with strength $J$ proportional to the Frank elastic constant. The second term represents the interaction between a liquid-crystal molecule at site $i$ and polymer unit at adjacent site $j$, with anchoring strength $H$. The third term is the effect of the applied electric field on the liquid crystal. We simulate this Hamiltonian using the Metropolis algorithm to visualize the director configuration and determine the orientational order as a function of temperature and applied field.

\begin{figure}
\includegraphics[width=3.35in]{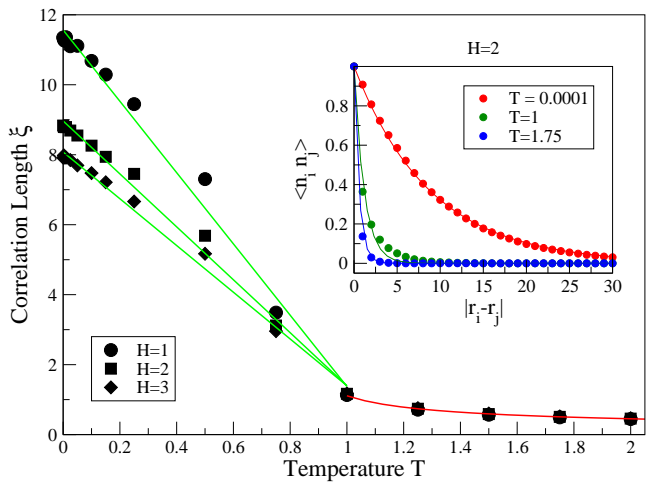}
\caption{Dependence of correlation length on temperature for various anchoring strengths: points are results from computer simulations, red and green lines are fits for high and low temperatures, respectively. Inset: Correlation functions for various temperatures, fit to an exponential form.}
\label{fig_sim_cor_length_vs_temp}
\end{figure}
Figure~\ref{fig_sim_temp} shows snapshots of the system. In the simulation, the system begins in a disordered state at high temperature, and undergoes a transition to a nematic state as the temperature is lowered. In the low-temperature final state, it consists of uniform domains of different director orientations, which form due to the competition between elastic energy favoring alignment ($J$ term) and the quenched disorder favoring randomness ($H$ term). If there were no polymer, the system would form one domain of uniform orientation.

To characterize the size of the domains, we measure correlations between director orientations. We compute the correlation function $g(|\mathbf{r}-\mathbf{r'}|) = \langle 2(\mathbf{n}(\mathbf{r}) \cdot \mathbf{n}(\mathbf{r'}))^2 - 1 \rangle$
as a function of distance, as shown in the Fig.~\ref{fig_sim_cor_length_vs_temp} inset for various temperatures. The results are fit to the exponential $g(|\mathbf{r}-\mathbf{r'}|) = \exp\left(-|\mathbf{r}-\mathbf{r'}|/\xi\right)$ to obtain the correlation length $\xi$. Figure~\ref{fig_sim_cor_length_vs_temp} shows the simulation data for correlation length as a function of temperature for various anchoring strengths; fits are discussed below.

\begin{figure}
\includegraphics[width=3.375in]{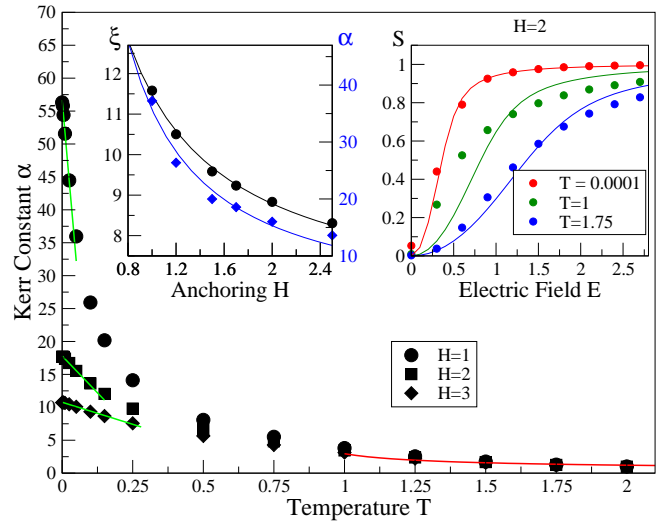}
\caption{Dependence of the Kerr constant $\alpha$ on temperature for various anchoring strengths. Left inset: Dependence of the correlation length $\xi$ and Kerr constant $\alpha$ on polymer anchoring energy $H$ in the limit of low temperature.  Right inset: Dependence of the scalar order parameter $S$ on external electric field $E$, for various temperatures, fit to the functional form given in Eq.~(\ref{S_prediction}).}
\label{fig_sim_kerr_const_vs_temp}
\end{figure}

Next, we apply an external electric field. Figure~\ref{fig_sim_temp}(b) shows snapshots of the director configurations for weak and strong electric fields, respectively. For a sufficiently strong field, all the liquid-crystal domains become fully aligned with the direction of the applied field. For weaker fields, the domains are not fully aligned, but still have some net order along the field direction. To quantify the orientational order, we anneal the system to the desired temperature, then run at that temperature to calculate the tensor order parameter $Q_{\alpha\beta}= \langle 2 n_\alpha n_\beta - \delta_{\alpha\beta} \rangle,$
where $\alpha$, $\beta = 1$, $2$, and the average is taken over \emph{both} sites and Monte Carlo steps. The positive eigenvalue of that average tensor yields the scalar order parameter $S$. We then apply the electric field, wait for the system to equilibrate, and measure the order parameter again. The dependence of the order parameter on applied electric field is shown by the data points in Fig.~\ref{fig_sim_kerr_const_vs_temp}, right inset.

To analyze the simulation results for $S(E)$, we would like to fit the data to a function and extract a parameter to characterize how sensitively the order responds to the field. To construct a fitting function, we suppose that each domain responds as a single unit to the applied field, with all spins in the domain at an angle $\theta$ relative to the field direction. The effective free energy of the domain should be
$F/k_B T = -\alpha E^2 \cos^2 \theta$.
The probability distribution for $\theta$ then becomes
$p(\theta) = e^{\alpha E^2 \cos^2\theta}/\int_0^{2 \pi} d\theta e^{\alpha E^2 \cos^2\theta}$,
and the average order parameter becomes
\begin{equation}
S(E)=\langle \cos 2\theta\rangle = \frac{\int_0^{2 \pi} d\theta \cos 2\theta e^{\alpha E^2 \cos^2\theta}}{\int_0^{2 \pi} d\theta e^{\alpha E^2 \cos^2\theta}}=\frac{J_1(\alpha E^2/2)}{J_0(\alpha E^2/2)}.
\label{S_prediction}
\end{equation}
We can now fit the functional form of Eq.~(\ref{S_prediction}) to the simulation results for $S(E)$, treating $\alpha$ as an arbitrary fitting parameter. The fits are shown by the solid lines in Fig.~\ref{fig_sim_kerr_const_vs_temp}, right inset. They are fairly good, although certainly not perfect, because the domain argument is only an approximation.

Based on the fits, we can take the parameter $\alpha$ as our measure of the response of the material to the electric field, which can be compared with the experimental Kerr constant. The conventional experimental definition of the Kerr constant $B$ is $\Delta n = B \lambda E^2$, where $\Delta n$ is the induced optical birefringence and $\lambda$ the optical wavelength~\cite{yang11}. The optical birefringence is related to the orientational order parameter by $\Delta n=\Delta n_\mathrm{max} S$. By comparison, for a small field our functional form of Eq.~(\ref{S_prediction}) reduces to $S(E) = \alpha E^2 / 4$. The conventional Kerr constant is therefore $B = \alpha \Delta n_\mathrm{max} / (4 \lambda)$. Hence, $\alpha$ provides the statistical mechanical information in the conventional Kerr constant, but not the optical information of $\Delta n_\mathrm{max}$ and $\lambda$. For convenience, we refer to $\alpha$ as the Kerr constant for the rest of this paper.

Figure~\ref{fig_sim_kerr_const_vs_temp} shows the parameter $\alpha$ as a function of temperature for various anchoring strengths. This Kerr constant increases as the temperature is reduced. For low anchoring the dependence on temperature is quite strong, but for high anchoring the dependence is much weaker. This behavior must arise from the competition among two ordering influences (Frank elastic energy and electric field) and two disordering influences (anchoring to quenched polymer disorder and thermal fluctuations). The challenge is to understand how $\alpha$ depends on these competing influences, and how it is related to the correlation length shown in Fig.~\ref{fig_sim_cor_length_vs_temp}.

To understand the simulation results for the Kerr constant, we develop an approximate theory for the susceptibility of a liquid crystal embedded in a random polymer network to an applied electric field. We build on earlier work on cooperative chiral order in random copolymers~\cite{selinger96,green99}, in which there is a pseudoscalar order parameter defined along a 1D chain. Randomness arises from quenched disorder in the sequence of right- and left-handed chemical groups, as well as from thermal fluctuations, and a net order parameter is induced by a slight population difference between the right- and left-handed groups. Here, we have a tensor order parameter defined on a 2D lattice (in the simulations) or 3D fluid (in the experiments). Randomness arises again from both quenched disorder and thermal fluctuations, and a net order parameter is induced by a slight electric field.

Following the example of that polymer problem, we consider the issues in three steps. First, we analyze the behavior of the system at zero temperature and zero applied field, with quenched disorder competing with the elastic energy. In that case we can estimate the characteristic domain size or correlation length $\xi$ as a function of anchoring strength $H$. Second, we investigate the response of domains of size $\xi$ to the symmetry-breaking field, and determine the Kerr constant $\alpha$ as a function of $H$. Third, we add thermal fluctuations and see how they reduce the order compared with quenched disorder only.

One complication in our simulation is that the quenched randomness is concentrated on specific sites corresponding to the polymer network. As an approximation, we map the model onto a system in which there is a random field on every site, as in a random-field magnetic system~\cite{imry75}. This approximation should be especially reasonable if each domain contains many polymers, i.~e.\ if the domain size is much greater than the typical polymer spacing. At each point of space we have a liquid-crystal director $\mathbf{n}(\mathbf{r})$ (in continuum notation) or $\mathbf{n}_i$ (in the discretized approach), and a local random field $\mathbf{h}(\mathbf{r})$ or $\mathbf{h}_i$. The energy has two terms: an elastic energy favoring alignment of neighboring directors
\begin{eqnarray}
\mathscr{H}_\mathrm{elastic} &=& \frac{K}{2}\int d\mathbf{r} [\partial_\alpha \mathbf{n}_\beta(\mathbf{r})][\partial_\alpha \mathbf{n}_\beta(\mathbf{r})] \nonumber\\
&=& J \sum_{\langle i,j \rangle}{\left[ 1 - \left(\mathbf{\hat{n}_i} \cdot \mathbf{\hat{n}_j}\right)^2 \right]}
\label{hamiltonian_cImry_Ma}
\end{eqnarray}
and a random term favoring alignment of the director with the local random field
\begin{eqnarray}
\mathscr{H}_\mathrm{random} = -\int d\mathbf{r}[\mathbf{h}(\mathbf{r})\cdot\mathbf{n}(\mathbf{r})]^2  = -H \sum_i ( \mathbf{\hat{h}}_i \cdot\mathbf{\hat{n}}_i )^2.
\label{hamiltonian_cImry_Ma_2}
\end{eqnarray}
Here, $K$ is the Frank elastic constant, and $J$ the corresponding strength of interaction between neighboring sites. $\mathbf{h}$ is a vector with magnitude $h=\sqrt{H}$ and random orientation $\mathbf{\hat{h}}_i$ at each site; the director $\mathbf{n}$ is a unit vector.

At zero temperature the system consists of domains, with uniform orientation inside each domain and negligible correlations among different domains (see Fig.~\ref{fig_sim_temp}(b)). We estimate the typical domain size $\xi$ using the classic Imry-Ma argument~\cite{imry75}. Domains form when the elastic energy (favoring large domains) is comparable to the random-field energy (favoring small domains). In spatial dimension $d$, with a continuous (not Ising) order parameter, the elastic energy scales as $\mathscr{H}_\mathrm{elastic} \sim K \xi^{d-2}$, while the random-field energy scales as $\mathscr{H}_\mathrm{random} \sim H \xi^{d/2}$. Equating these energies gives a prediction for the domain size
\begin{equation}
\xi \sim \left(K/H\right)^{2/(4-d)}.
\label{cor_without}
\end{equation}
Note that this domain size is large when the random fields are small, and decreases as the random fields increase.

At this point, we should take into account one difference between the random-field system and the disordered polymer network. In the random-field system, the domain size can become arbitrarily small as the field strength increases. By contrast, in the disordered polymer network, the domain size can never become smaller than the characteristic spacing of the polymer network, denoted $\xi_{\infty}$, even if the anchoring strength $H\to\infty$. Hence, we modify Eq.~(\ref{cor_without}) to show this minimum domain size
\begin{equation}
\xi \sim \left(K/H\right)^{2/(4-d)}+\xi_{\infty}.
\label{cor_with}
\end{equation}

In Fig.~\ref{fig_sim_kerr_const_vs_temp}, left inset, the black circles show simulation results for the correlation length as a function of anchoring in the low temperature limit. By comparison, the black line is Eq.~(\ref{cor_with}) in $d=2$, as appropriate for the simulations, $\xi \sim \frac{K}{H} + \xi_\infty$.
With the single fitting parameter $K/\xi_{\infty}$, we obtain a good fit to the simulation.

We calculate the Kerr constant by considering how a domain of size $\xi$ responds to a symmetry-breaking field, in the limit of zero temperature and averaging over a statistical distribution of random fields (see supplemental material). In the limit of small electric field, we obtain the Kerr constant
\begin{equation}
\alpha = (2\pi)^{1/2} \xi^{d/2}H^{-1}.
\label{kerr_small}
\end{equation}
Note that $\alpha$ is proportional to $\xi^{d/2}$, the square root of the domain volume, not to the domain volume itself. This square root dependence is a signature of statistical randomness of the quenched random fields as opposed to thermal randomness. An analogous square root dependence of the susceptibility was seen in the earlier study of chiral order in random copolymers~\cite{selinger96}.

We can combine Eq.~(\ref{kerr_small}) for the Kerr constant with Eq.~(\ref{cor_with}) for the correlation length to obtain
\begin{equation}
\alpha \sim H^{-1} \left[\left(K/H\right)^{2/(4-d)}+\xi_{\infty}\right]^{d/2}.
\label{kerr_with}
\end{equation}
Note that $\alpha$ increases as the elastic constant $K$ increases, and decreases as the anchoring $H$ to the disordered polymer network increases.

In Fig.~\ref{fig_sim_kerr_const_vs_temp}, left inset, the blue points show the simulation results for the Kerr constant as a function of anchoring strength in the low temperature limit. By comparison, the blue line shows the prediction of Eq.~(\ref{kerr_with}) in $d=2$, as appropriate for the simulations, $\alpha \sim K/H^2 + \xi_{\infty}/H.$ As in the previous section, we obtain a good fit to the simulation with the single fitting parameter $K/\xi_{\infty}$.

For the dependence of correlation length on temperature, there are distinct limiting cases at high temperature and low temperature.  At high temperature, the free energy density takes the form
\begin{equation}
F = \frac{k_B (T-T^*)}{2A} Q_{\alpha\beta}Q_{\alpha\beta} + \frac{\kappa}{2} (\partial_\gamma Q_{\alpha\beta})(\partial_\gamma Q_{\alpha\beta}),
\end{equation}
where $A$ is the area per molecule and $\kappa$ the elastic constant for variations in the tensor order parameter. By comparing the two terms in $F$,
we can estimate the correlation length as
\begin{equation}
\xi \sim \sqrt{\frac{A\kappa}{ k_B (T-T^*)}}.
\end{equation}

At zero temperature, the correlation length is given by Eq.~(\ref{cor_without}). For low but nonzero temperature, $K$ generally varies with temperature as
$K(T) = K_0 - K'  T,$ where $K'$ is a phenomenological slope, leading to the correlation length
\begin{equation}
\xi(T,H) = \frac{K_0}{H} -\frac{K' T}{H} + \xi_\infty.
\label{eq_cor_vs_temp}
\end{equation}

In Figs.~\ref{fig_sim_cor_length_vs_temp} and~\ref{fig_sim_kerr_const_vs_temp} we fit our simulation results with the predictions in low (green) and high (red) temparature regimes. For the correlation length, we obtain a good fit over the entire range of temperature, but for the Kerr constant the fit works only in the limits of low and high temperature. For intermediate temperature, the competion of the energies is complex and is not described well by the approximations in this section, so we can only make predictions using the simulations.

In this paper, we have performed computer simulations and developed an approximate analytic theory to describe nematic liquid crystals embedded in disordered polymer networks. The main result of this work is that the low-temperature properties are controlled by the statistical randomness of the polymer network, rather than by thermal fluctuations. At low temperature, deep in the nematic phase, the local orientational order is well-developed and the key issue is the long-range domain structure. We find that the domain size is given by the Imry-Ma argument, which balances the elastic energy of director variations against the anchoring energy associated with the local random polymer directions. Likewise, the Kerr constant is given by the response of uniform domains to the statistical distribution of the local field, which is slightly biased by the applied electric field. The Kerr constant is large because the local orientational order is well-developed, and it is fairly temperature-independent because it is
mainly controlled by the randomness in the polymer network. These predictions may be used in the development of materials for electro-optics, thus making a connection between the fundamental physics of liquid crystals in quenched disorder and technological applications.

We would like to thank D.-K. Yang for many helpful discussions. This work was supported by the National Science Foundation through Grant DMR-1106014.

\onecolumngrid
\newpage
\section*{Supplemental Material}
\twocolumngrid
\setcounter{equation}{0}
\setcounter{figure}{0}

Here we present the details of the derivation of the Kerr constant. We consider how a domain of size $\xi$ responds to a symmetry-breaking field, in the limit of zero temperature. The energy of the domain is the sum of the applied-field energy and the random-field energy
\begin{equation}
\mathscr{H} = -\sum_i \left[ (\mathbf{E} \cdot \mathbf{\hat{n}})^2 + H(\mathbf{\hat{h}}_i \cdot \mathbf{\hat{n}})^2 \right],
\label{hamiltonian_cs_2}
\end{equation}
where $\mathbf{\hat{n}}$ is the uniform director inside the domain.

Let us choose the $x$-direction to be the direction of the applied field, write the random field orientation at site $i$ as $\mathbf{\hat{h}}_i = (\cos \varphi_i,\sin \varphi_i)$, and write the director as $\mathbf{\hat{n}}=(\cos\theta,\sin\theta)$. The domain energy then becomes
\begin{equation}
\mathscr{H} = \mathrm{const} - \frac{\xi^d E^2 + H h'}{2} \cos 2\theta - \frac{H h''}{2} \sin 2\theta,
\label{hamiltonian_cs_4}
\end{equation}
where $h'=\sum_i \cos 2 \varphi_i$ and $h''=\sum_i \sin 2 \varphi_i$. At zero temperature, the director goes to its lowest-energy state, which is
\begin{eqnarray}
\label{cos_sin}
\cos 2\theta &=& \frac{\xi^d E^2 + H h'}{\sqrt{(\xi^d E^2 + H h')^2 + (H h'')^2}},\\
\sin 2\theta &=& \frac{H h''}{\sqrt{(\xi^d E^2 + H h')^2 + (H h'')^2}}.\nonumber
\end{eqnarray}
Hence, the nematic order tensor for the domain is
\begin{eqnarray}
\label{q_single_realization}
Q^\mathrm{domain}_{\alpha\beta}=
\begin{pmatrix}
\cos 2\theta & \sin 2\theta\\
\sin 2\theta & -\cos 2\theta
\end{pmatrix}
\end{eqnarray}
for any particular realization of the random fields.

We must now average over the statistical distribution of random fields. The probability distribution function for the sums $h'$ and $h''$ is the Gaussian
\begin{equation}
P(h',h'') = \frac{1}{\pi \xi^d} e^{-(h'^2 + h''^2)/(2 \xi^d)}.
\label{dist_P}
\end{equation}
The average of the nematic order tensor over this distribution can be written as
\begin{eqnarray}
\label{q_average}
Q_{\alpha\beta} = \langle Q^\mathrm{domain}_{\alpha\beta} \rangle =
\begin{pmatrix}
S & 0\\
0 & -S
\end{pmatrix},
\end{eqnarray}
where
\begin{equation}
S = \langle \cos 2 \theta \rangle = \int dh' dh'' \frac{(\xi^d E^2 +H h') P(h',h'')}{\sqrt{(\xi^d E^2 +H h')^2 + (H h'')^2}}.
\label{ave_cos}
\end{equation}
We expand Eq.~(\ref{ave_cos}) for small electric fields, and after integration obtain $S = \alpha E^2 /4$ with Kerr constant
\begin{equation}
\alpha = \frac{(2\pi)^{1/2} \xi^{d/2}}{H}.
\end{equation}

\end{document}